\documentclass[doublecol,a4paper,showpacs]{epl2}
\usepackage{tensor}
\usepackage{graphicx}
\usepackage{amsmath}
\usepackage{amssymb}
\usepackage{enumerate}
\usepackage{subfigure}
\usepackage{tabularx}
\usepackage[colorlinks=true, pdfstartview=FitV, linkcolor=blue, citecolor=red, urlcolor=black, breaklinks=true]{hyperref}
\newcommand{\be}{\begin{equation}}
\newcommand{\ee}{\end{equation}}
\newcommand{\ben}{\begin{eqnarray}}
\newcommand{\een}{\end{eqnarray}}
\newcommand{\bes}{\begin{subequations}}
\newcommand{\ees}{\end{subequations}}
\def\bal#1\eal{\begin{align}#1\end{align}}

\newcommand{\bb}{\bibitem}
\newcommand{\sech}{{\rm sech}}
\newcommand{\LL}{{\mathcal L}}

\title{Novel lumplike structures}
\author{M.A. Marques\inst{1,2}\thanks{\email{mam.matheus@gmail.com}}}
\shortauthor{M.A. Marques \etal}

\institute{ \inst{1} Department of Physics and Astronomy, University of Pennsylvania, Philadelphia, PA 19104, USA \\                
  \inst{2} Departamento de F\'\i sica, Universidade Federal da Para\'\i ba, 58051-970 Jo\~ao Pessoa, PB, Brazil}


\abstract{We study the existence of new features in lumplike solutions in models of a real scalar field in two dimensional flat spacetime. We present new models and field configurations that exhibit a non standard decay, shrinking or stretching the tail of the solutions.}

\date{\today}
\begin{document}
\maketitle

\section{Introduction}
In high energy physics, defect structures can be of a topological or non-topological nature. Over the years, they have been investigated severaly  \cite{wilets,vilenkin,manton} and may find applications in other areas in Physics \cite{davidov,agrawal,fradkin}. The simplest topological structure that appear in classical field theory are called kinks \cite{vachaspati}. They arise in $(1,1)$ flat spacetime dimensions in an action of a single real scalar field and are linearly stable. Regarding the non topological objects, the simplest ones are lumps, which appear in a similar action. The linear stability of the lumps, and similarly for the kinks, is investigated through a Schr\"odinger-like equation, which shows, in some manner, a connection with quantum mechanics \cite{bazeiaolmo}. Lumps are unstable under small fluctuations. However, their instability may be of use in the modeling of dissipation in quantum gravity \cite{inst1} and classical mechanics\cite{inst2}.

The instability of localized structures may be worked around by enlarging the model to find mechanisms to stabilize them. So, they are not ruled out of Physics. For instance, in Refs.~\cite{macpherson,bm}, one have dealt with the case of a fermionic ball, which may appear with the inclusion of charged fermions in a manner that they become entrapped inside the collapsing solution, charging it and stabilizing the whole structure. Other possibility is to treat the scalar field as an axion field \cite{a1,a2,a3}.

We then focus on lumplike structures, which are of interest in many areas of Physics. For instance, in high energy physics, they may be seeds for the formation of structures \cite{frieman,macpherson,coulson,khlopov,bm}. Furthermore, they may be useful in the investigation of other structures, such as axions \cite{R}, Q-balls \cite{coleman,qbreview,qbsplit} and skyrmions \cite{sk1,sk2,sk3}, or in models of inflation \cite{LR} and braneworlds \cite{B,BJ}. In condensed matter physics, they may be associated to the transport of charge in diatomic chains \cite{pnevmatikos,xu1,xu2,bnt,bllm} and are important in optic fibers \cite{agrawal,haus}, where they can represent bright solitons.

In this paper, we follow a similar direction to the study implemented in \cite{ave,wes}. Here, however, we focus on a different issue, dealing with lumplike solutions with new features related to their decay. This study is also motivated by Refs.~\cite{high,fkthc,polytail}, which present kinks with non standard decay. The work is organized as follows: in the next section, we present the general model and its properties, including the usual behavior of the decay of lumplike solutions. We continue the investigation revisiting the standard lump and presenting two new models that engender novel features. We end our work presenting conclusions and perspectives for future investigations.

\section{Generalities}
We work in the flat spacetime scenario with the metric tensor given by $\eta_{\mu\nu}=\textrm{diag}(1,-1)$. To investigate models that allow for the presence of lumps, we consider the action $S=\int d^2x \LL$, with Lagrangian density in the form
\be\label{lmodel}
{\cal L} = \frac12 \partial_\mu\phi \partial^\mu \phi - V(\phi),
\ee
where $V(\phi)$ denotes the potential. For simplicity, we work with dimensionless quantities. One may vary the action associated to the above Lagrangian density with respect to $\phi$ to obtain the field equation
\be\label{eomt}
\ddot{\phi}-\phi^{\prime\prime} + V_\phi=0,
\ee
where the dot and the prime represent the derivative with respect to the time, $t$, and to the spatial coordinate, $x$, respectively, and also $V_\phi=dV/d\phi$. In order  to search for lumplike structures, we take static configurations. By doing so, the above field equation becomes
\be\label{eom}
\phi^{\prime\prime} = V_\phi,
\ee
which must be solved with the boundary conditions $\phi(x\to\infty) = \phi(x\to-\infty) = v$, where $v$ is a minimum of the potential. One can show that, in this case, the topological current $j^\mu = \epsilon^{\mu\nu} \partial_\nu \phi$ leads to a null topological charge, which makes the lumps be non-topological solutions. Before going further, we can estimate the asymptotic behavior of a lump by considering $\phi(x\to\infty) \to v + \phi^\pm_{asy}$ in the above equation. We then get
\be\label{asy}
\phi^\pm_{asy} \propto e^{\mp\sqrt{V_{\phi\phi}(v)}\,x}.
\ee
From the above equation, we see that $V_{\phi\phi}(v)$ controls the size of the tail of the solution.. Since this term is usually a positive finite constant, standard lumps have exponential decay. A different behavior may appear, for instance, if one introduces a parameter that controls the aforementioned quantity; this was done in Ref.~\cite{complump} as a route to compactify the lump.

The energy density can be calculated as usual; it has he form
\be\label{rho}
\rho(x)=\frac12 {\phi^\prime}^2 + V(\phi).
\ee

We now study the linear stability of the lump, i.e., we see how it behaves under small fluctuations, by considering $\phi(x,t) = \phi(x) + \sum_i \eta_i(x) \cos(\omega_i t)$, where $\phi(x)$ is the solution of Eq.~\eqref{eom}. We substitute this expression into the time-dependent field equation \eqref{eomt} and expand it up to first order in $\eta$ to get the stability equation
\be\label{sta}
-\eta_i^{\prime\prime} + U(x) \eta_i = \omega_i^2 \eta_i,\quad\text{where}\quad U(x) = \left. V_{\phi\phi} \right|_{\phi=\phi(x)},
\ee
represents the stability potential. The zero mode, $\eta_0(x)$, is given by
\be
\eta_0(x) = \phi^\prime(x).
\ee
The solution is unstable if the stability equation \eqref{sta} admits at least one negative $\omega_i^2$. Regarding lumps, the zero mode usually presents a node. This means that the stability equation supports a lower bound state with negative energy. Thus, lumps are unstable under small fluctuations.

\section{Examples}
Next, we review a model that exhibits lumps with the standard asymptotic behavior in Eq.~\eqref{asy}, and present two new models which engender lumps with novel features on their decay.
\subsection{Standard lump}
\begin{figure}[t!]
\centering
\includegraphics[width=4.2cm]{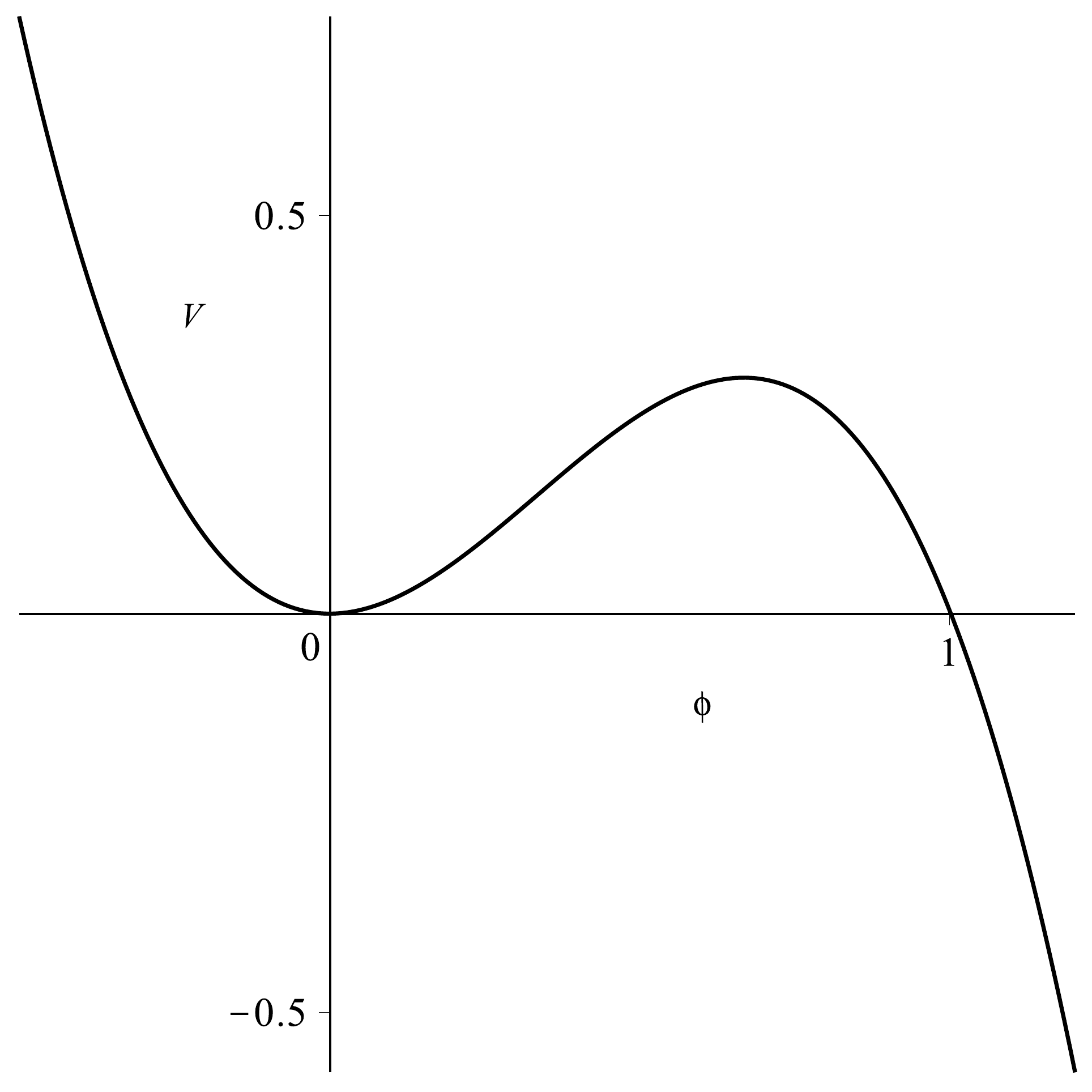}
\includegraphics[width=4.2cm]{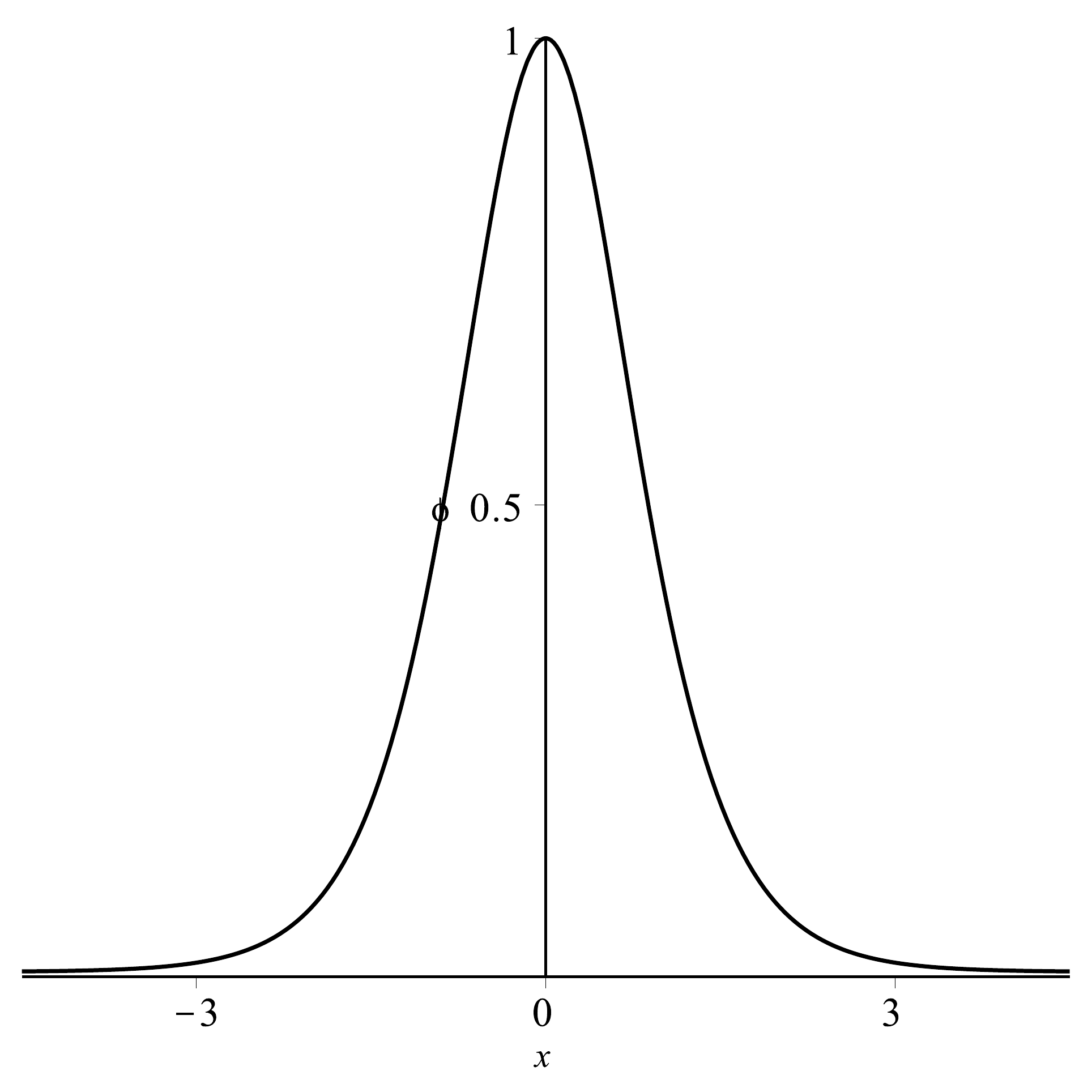}
\includegraphics[width=4.2cm]{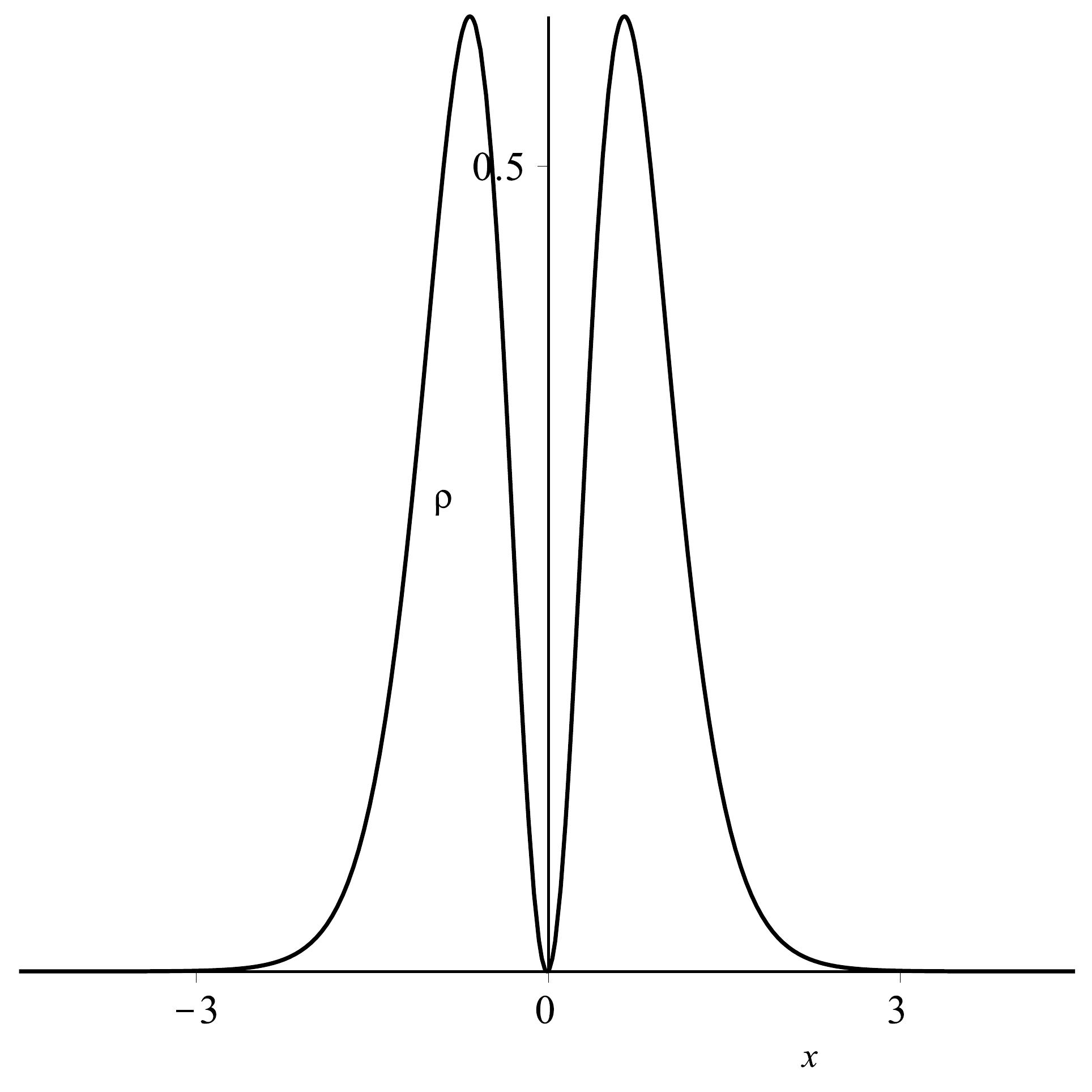}
\includegraphics[width=4.2cm]{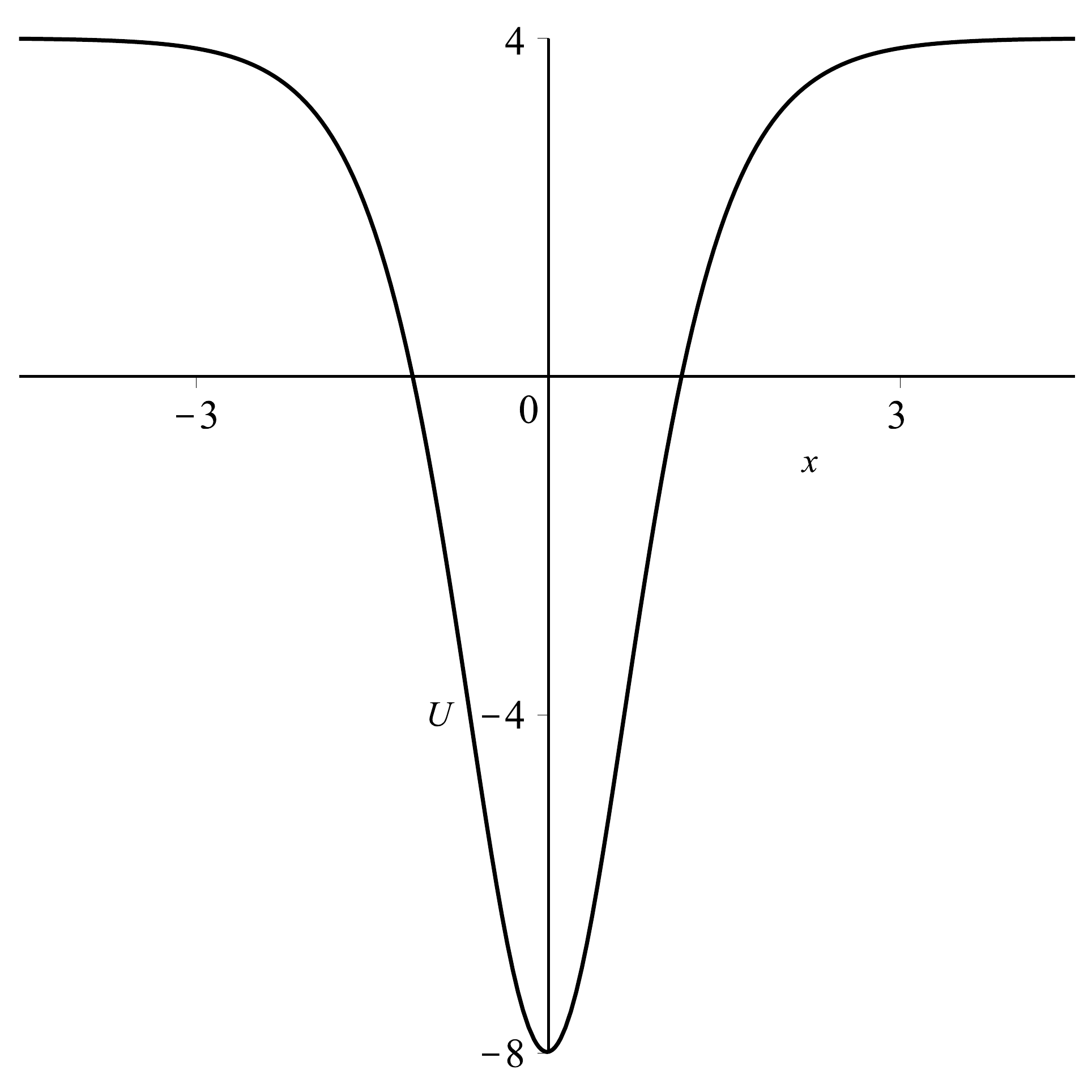}
\caption{The potential \eqref{vphi3} (top left), the solution \eqref{solphi3} (top right), the energy density \eqref{rhophi3} (bottom left) and the stability potential \eqref{uphi3} (bottom right) for the $\phi^3$ model.}
\label{fig1}
\end{figure}

To review the case of a standard lump, we consider the $\phi^3$ potential
\be\label{vphi3}
V(\phi)=2\phi^2(1-\phi),
\ee
which has a local minimum at $\phi=0$, with $V_{\phi\phi}(0)=4$, and a zero at $\phi=1$. It is displayed in Fig.~\ref{fig1}. The field equation \eqref{eom} takes the form
\be
\phi^{\prime\prime} = 2\phi(2-3\phi).
\ee
It admits the solution
\be\label{solphi3}
\phi(x) = \sech^2(x),
\ee
which is a lump, as one can see in Fig.~\ref{fig1} with a tail that decays exponentially. Its energy density can be calculated from Eq.~\eqref{rho}, which becomes
\be\label{rhophi3}
\rho(x) = 4\,\sech^4(x)\tanh^2(x),
\ee
and it is shown in Fig.~\ref{fig1}. By integrating it, one gets the energy $E=4/3$. The stability potential is given by Eq.~\eqref{sta}:
\be\label{uphi3}
U(x) = 4 - 12\,\sech^2(x).
\ee
It can be seen in Fig.~\ref{fig1}. Its respective stability equation supports the zero mode $\eta(x)=\sech^2(x)\tanh(x)$, which has a node. Thus, there is a negative bound state below it, which shows the instability of the $\phi^3$ lump under small fluctuations.

\subsection{Polynomial decay}
Structures with polynomial decay find applications in a wide range of scenarios. One can find them, for instance, by dealing with the statistical mechanics of solvable models that engender long range interactions \cite{poly1}, by investigating of dipolar quantum gases \cite{poly2} and by studying quantum information with Rydberg atoms \cite{poly3}. Usually, the forces that mediate the long range interactions in the aforementioned scenarios include the dipole-dipole and the van der Waals cases.

\begin{figure}[t!]
\centering
\includegraphics[width=6cm]{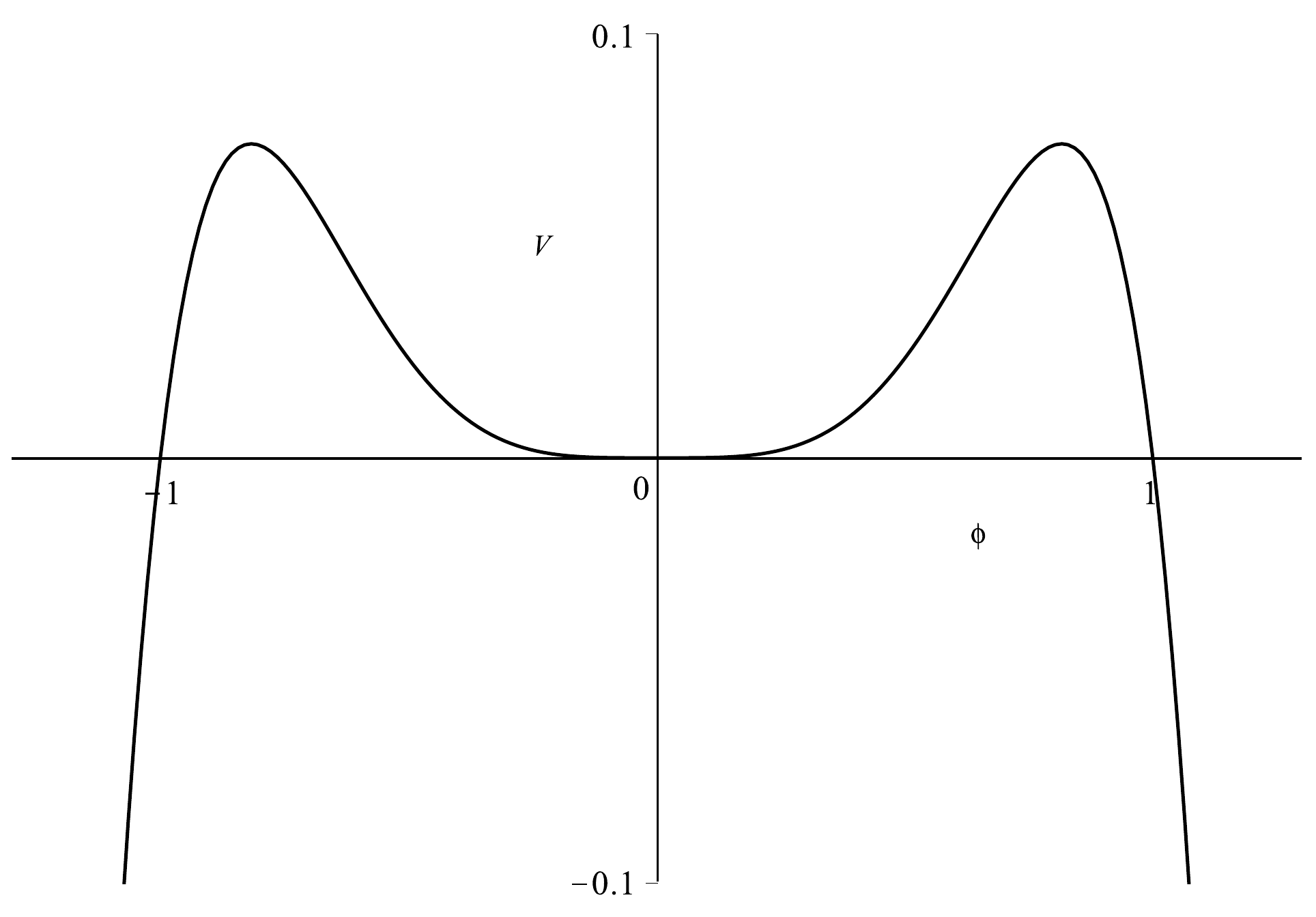}
\caption{The potential in Eq.~\eqref{vpoly}, which presents a large valley around the local minimum.}
\label{fig2}
\end{figure}

In this example, we introduce a new model that engenders a non standard asymptotic behavior. It is given by the polynomial potential
\be\label{vpoly}
V(\phi) = \frac12\phi^4\left(1-\phi^2\right).
\ee
This potential has a local minimum at $\phi=0$, with $V_{\phi\phi}(0)=0$, and zeroes at $\phi=\pm1$. It is displayed in Fig.~\ref{fig2}. Notice that its valley around the minimum is larger than in the $\phi^3$ model, given by the potential in Eq.~\eqref{vphi3}.

\begin{figure}[htb!]
\centering
\includegraphics[width=4.2cm]{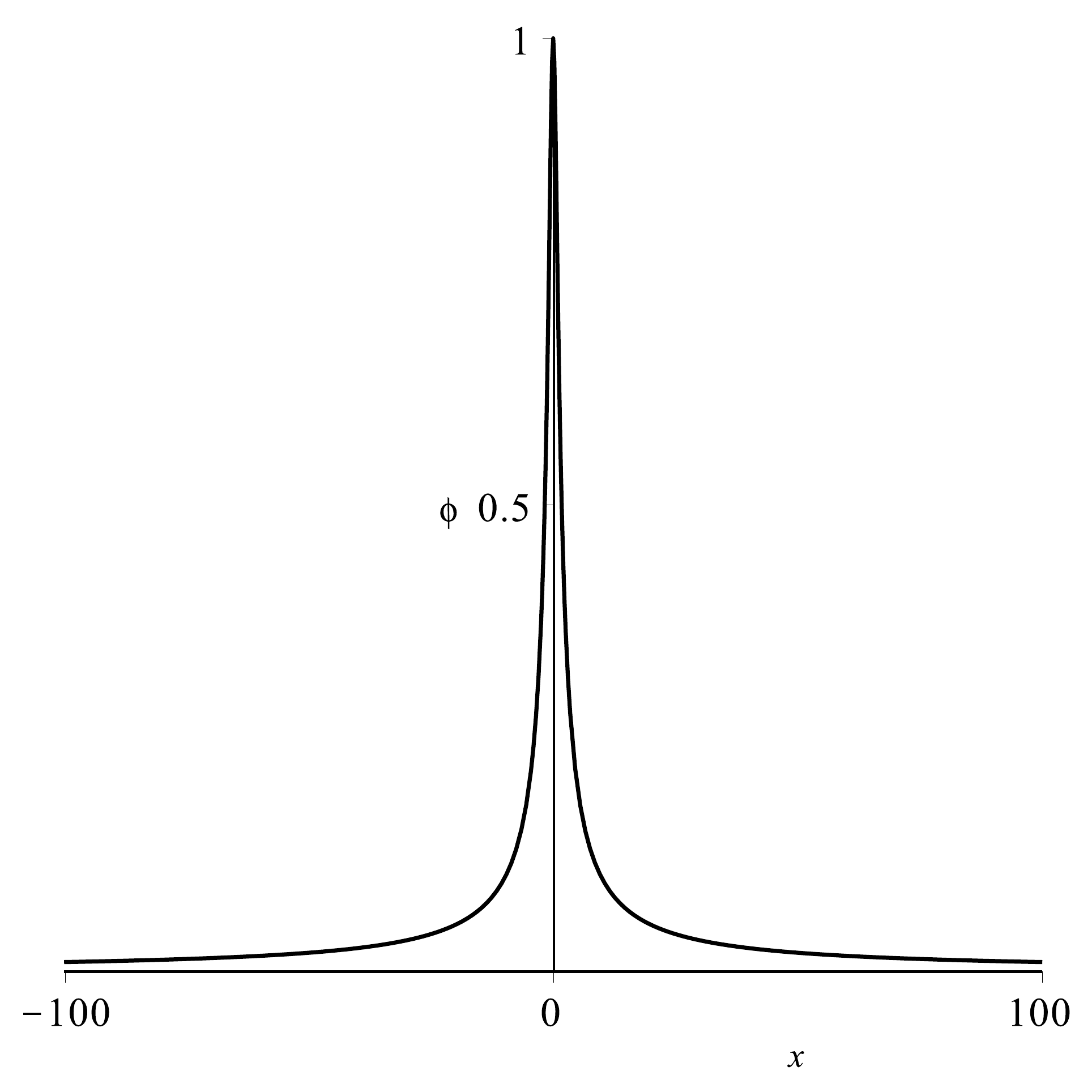}
\includegraphics[width=4.2cm]{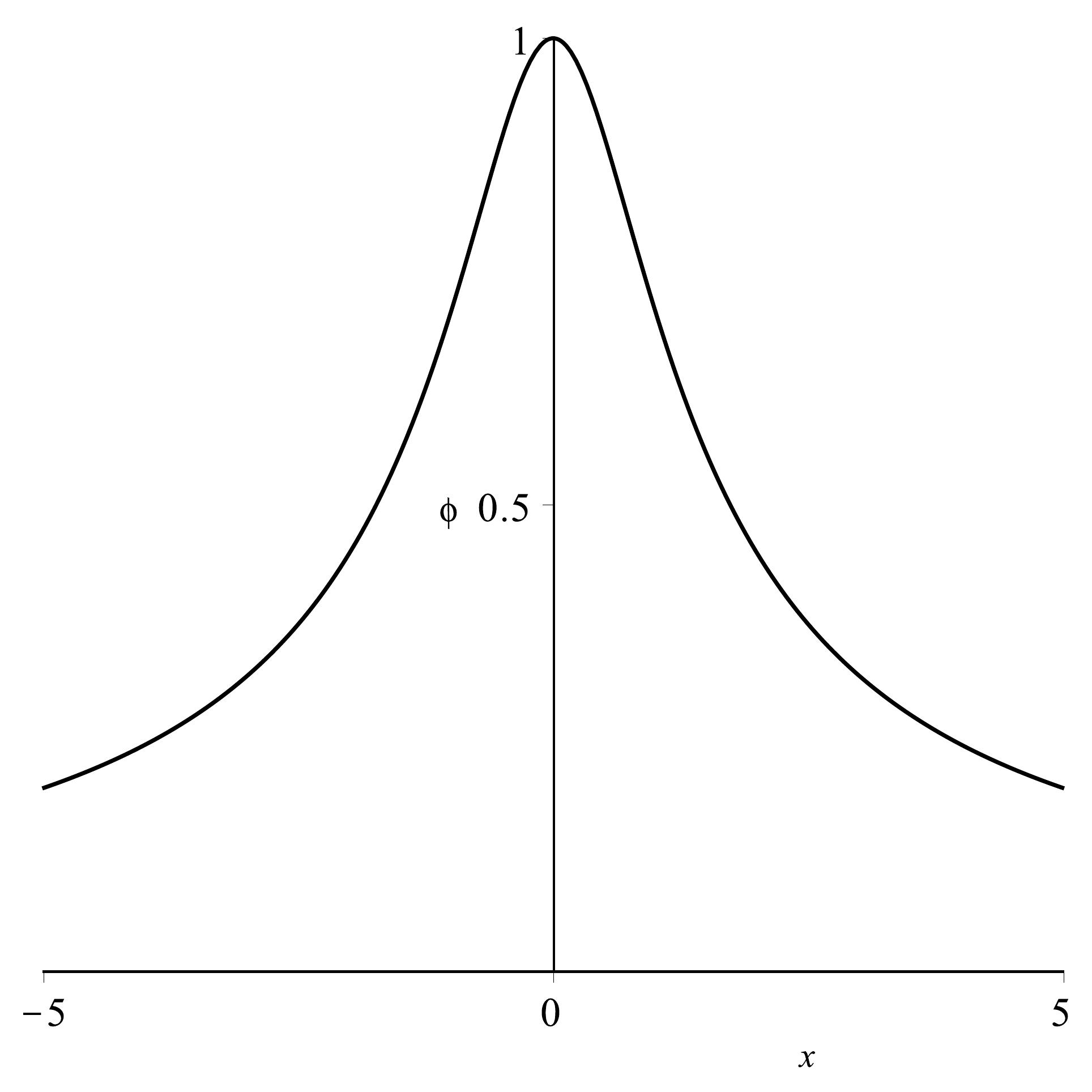}
\caption{The lumplike solution in Eq.~\eqref{solpoly}. In the left panel, we show its general behavior. In the right panel, we present its demeanor around the origin.}
\label{fig3}
\end{figure}

The field equation \eqref{eom} become
\be
\phi^{\prime\prime} = \phi^3\left(2-3\phi^2\right).
\ee
The above equation presents the symmetry $\phi\to-\phi$. So, we only consider $\phi\geq0$ and write the solution
\be\label{solpoly}
\phi(x) = \frac{1}{\sqrt{1+x^2}}.
\ee
Notice that its asymptotic behavior is $\phi(x\to\pm\infty) \to 1/|x|$. Therefore, we expect the solution to present a tail that goes farther than the one of the standard lump in Eq.~\eqref{solphi3} due to this slow polynomial decay. We see this feature in Fig.~\ref{fig3}, where we plot it.

\begin{figure}[t!]
\centering
\includegraphics[width=4.2cm]{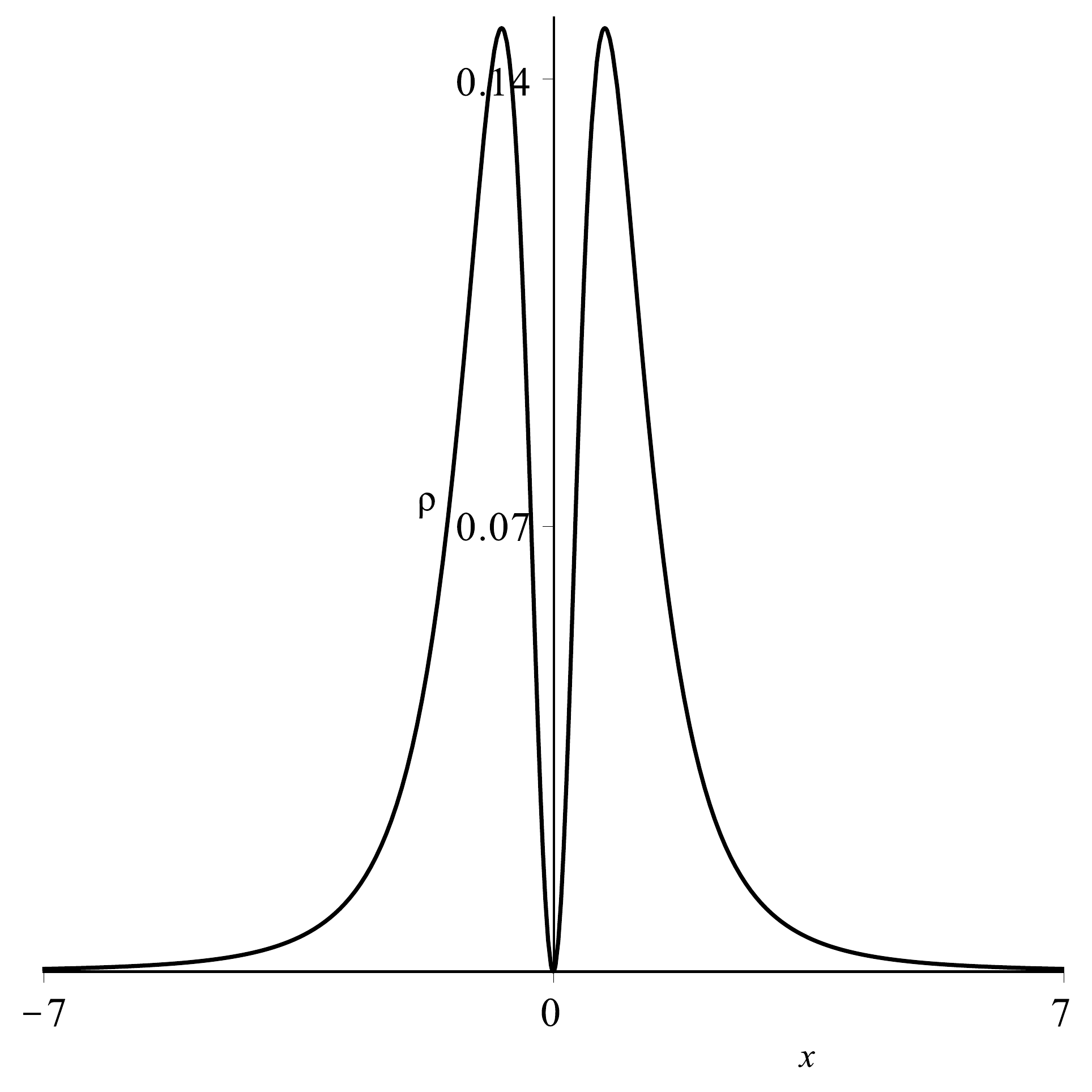}
\includegraphics[width=4.2cm]{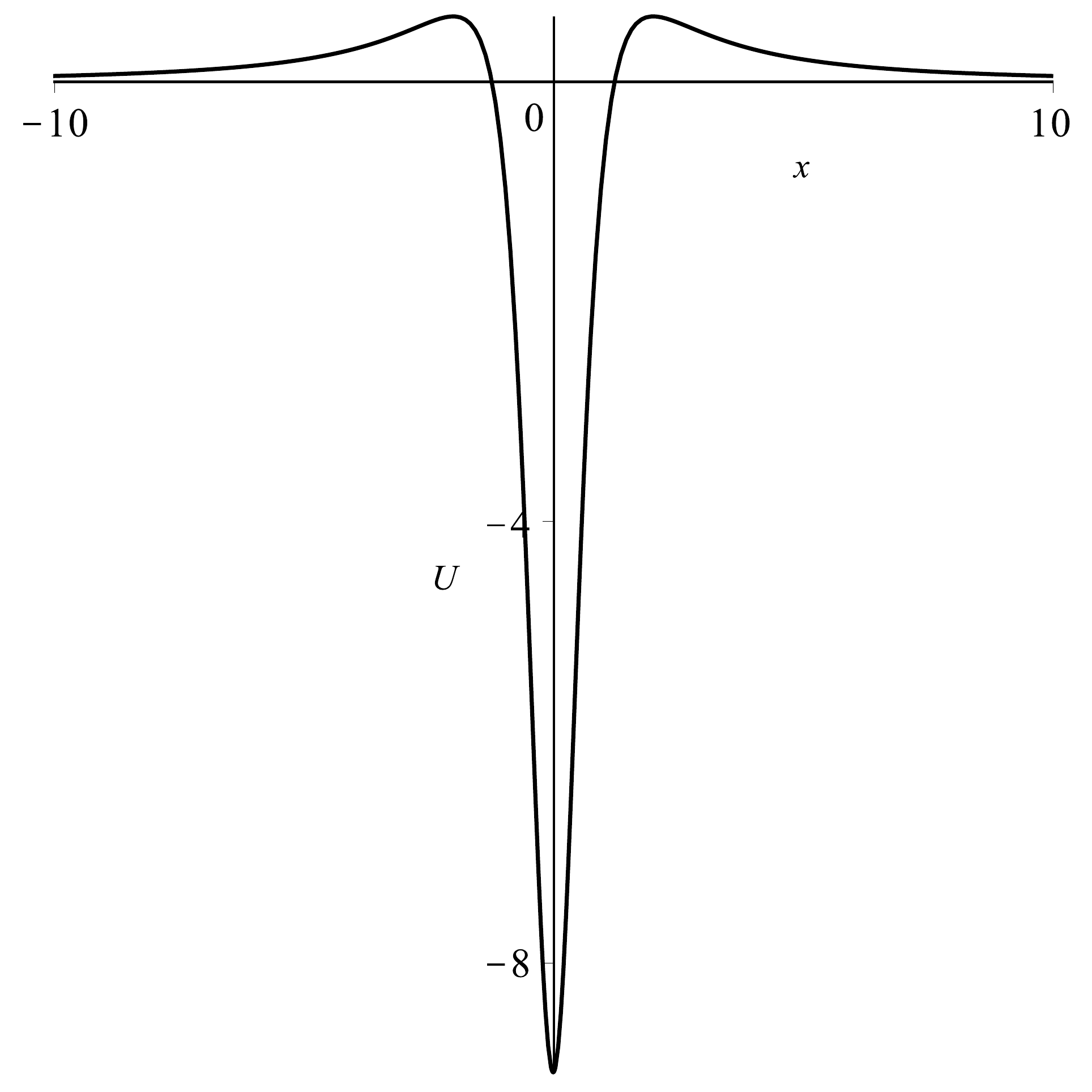}
\caption{The energy density in Eq.~\eqref{rhopoly} (left) and the stability potential in Eq.~\eqref{upoly} (right).}
\label{fig4}
\end{figure}

The energy density in Eq.~\eqref{rho} takes the form
\be\label{rhopoly}
\rho(x) = \frac{x^2}{\left(1+x^2\right)^3}.
\ee
It is displayed in Fig.~\ref{fig4}. We see the large tail of the solution is less evident in the energy density because of its asymptotic behavior, $\rho(x\to\pm\infty) \to 1/x^4$, which decays, indeed, faster than the solution. By integrating the above expression, we get the energy $E=\pi/8$.

The stability potential \eqref{sta} in this case is
\be\label{upoly}
U(x) = \frac{3\left(2x^2-3\right)}{\left(1+x^2\right)^2}.
\ee
In Fig.~\ref{fig4}, we plot it. Notice that $U(x\to\infty)\to0$, which happens due to the aforementioned feature of the potential minimum. It has a volcano profile and its associated zero mode is $\eta_0 = x/(1+x^2)^{3/2}$, which presents a node. Thus, the stability equation \eqref{sta} admits a negative eigenvalue, which makes the lump in Eq.~\eqref{solpoly} unstable under small fluctuations. 

It is worth commenting that the potential in Eq.~\eqref{vpoly} may be generalized to
\be
V(\phi) = \frac{2n^2}{m^2} \phi^{2+m/n}\left(1-\phi^m\right)^{2-1/n},
\ee
where $m$ is a positive even number and $n$ is a positive odd parameter. In this case, one gets the solution
\be\label{solpolyg}
\phi(x) = \frac{1}{\left(1+x^{2n}\right)^{1/m}}.
\ee
Notice that its asymptotic behavior is $\phi(x\to\pm\infty) \to 1/|x|^{2n/m}$. Thus, the parameters $m$ and $n$ controls how the lump decays. The energy density is
\be
\rho(x) = \frac{4n^2}{m^2}\frac{x^{4n-2}}{\left(1+x^{2n}\right)^{2+2/m}}.
\ee
One may also calculate the stability potential analitically, which is cumbersome, so we do not write it here. Since the zero mode is $\eta_0(x) = 2n x^{2n-1}/\left(m(1+x^{2n})^{1+1/m}\right)$, it presents a node. Thus, the lump in Eq.~\eqref{solpolyg} is unstable.

\subsection{Double exponential decay}
\begin{figure}[t!]
\centering
\includegraphics[width=4.2cm]{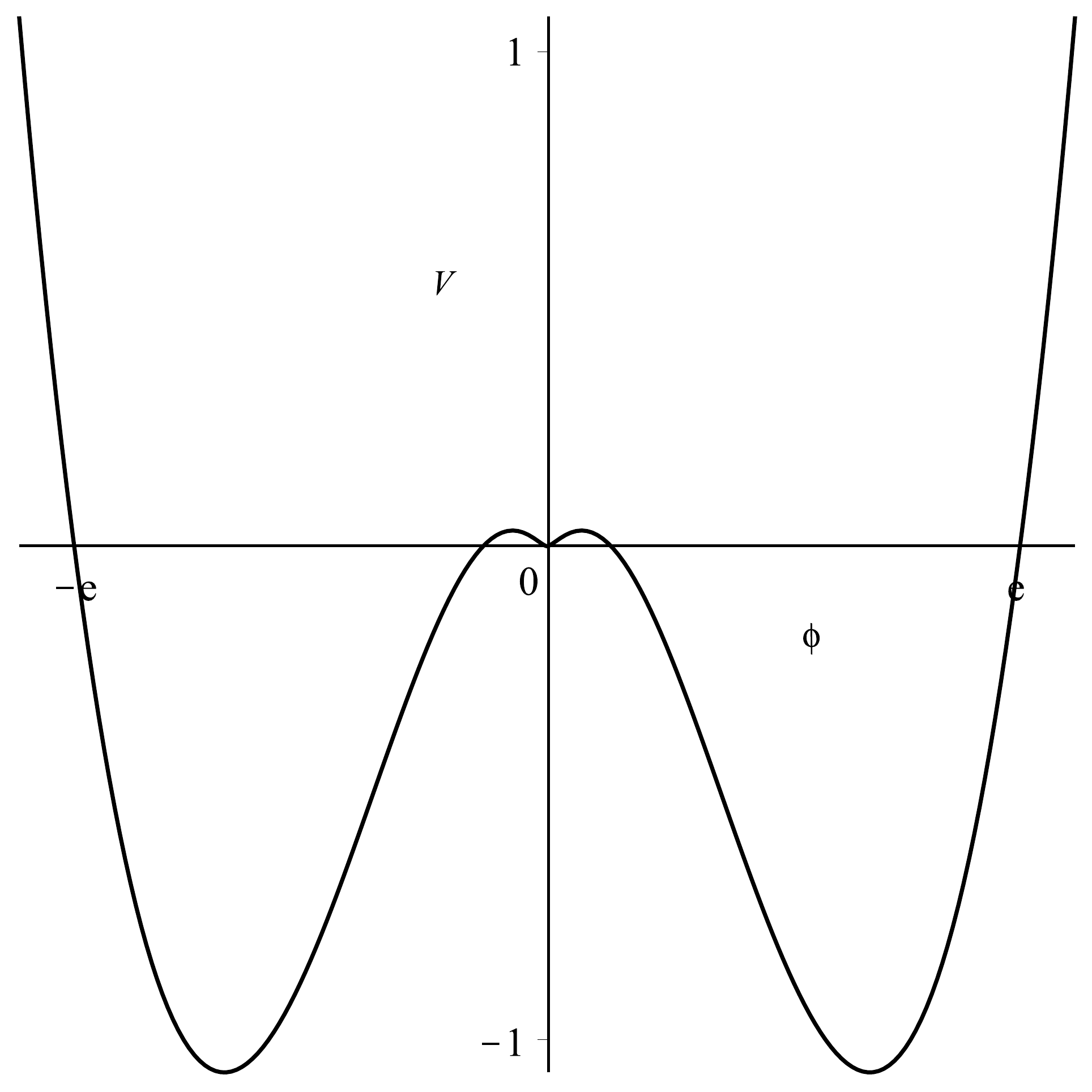}
\includegraphics[width=4.2cm]{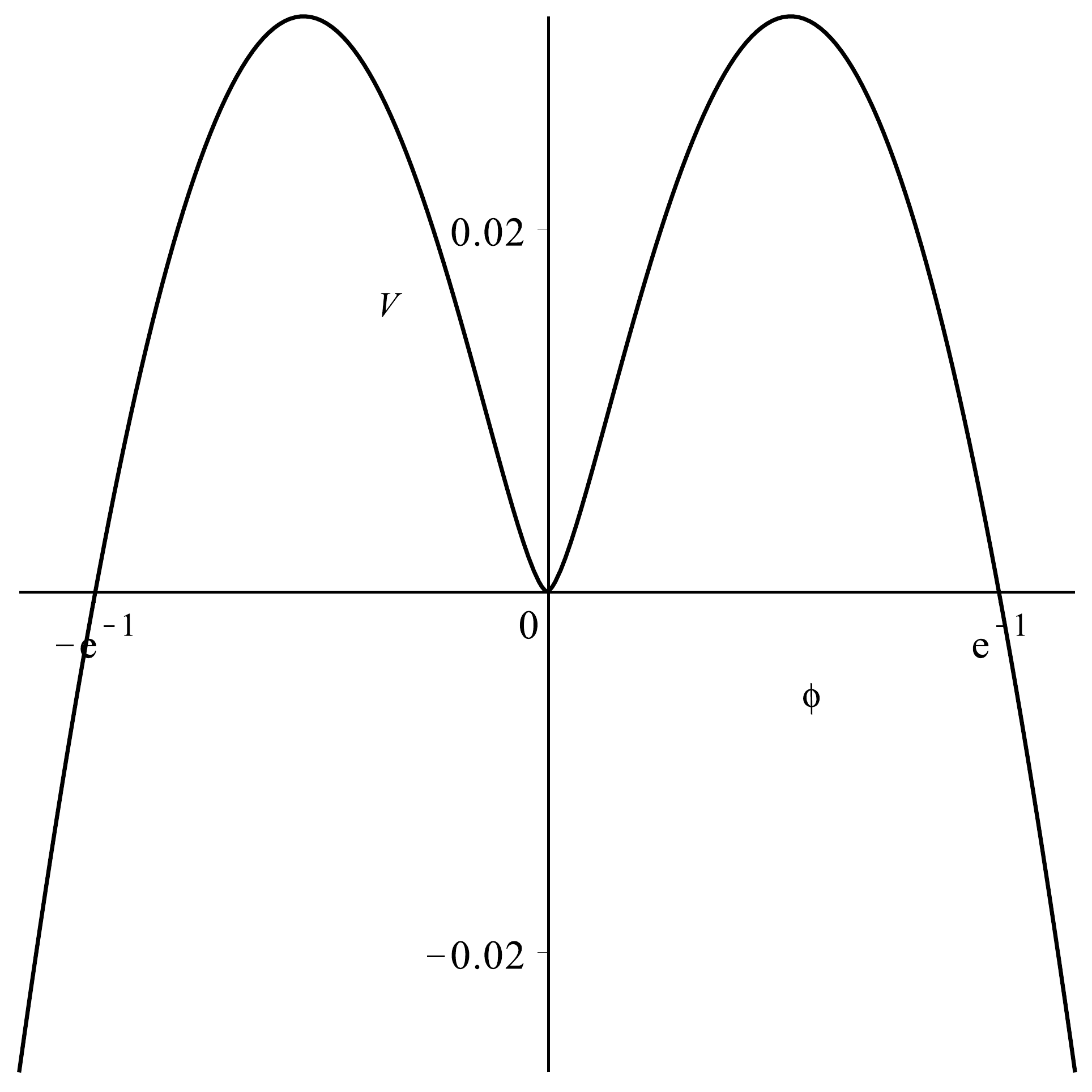}
\caption{The potential in Eq.~\eqref{vde}, which presents a large valley around the local minimum. In the left panel, we show its general behavior. In the right panel, we display its demeanor around the origin, where the lump that we are interested in exists.}
\label{fig5}
\end{figure}

Here, we present a model that support lumps with double exponential decay. It arises for the potential
\be\label{vde}
V(\phi) = \frac18\phi^2\!\left(\ln^2\!\left(\phi^2\right)-4\right).
\ee
In Fig.~\ref{fig5}, we plot the above potential. It has a local minimum at the origin, with $V(0)=0$, and two others at $\phi^\pm_{min}=\pm e^{(\sqrt{5}-1)/2}$ with $V(\phi^\pm_{min})=-e^{(\sqrt{5}-1)}(\sqrt{5}-1)/4$. The zeroes are at $\phi=\pm e^{-1}$ and $\phi=\pm e$. The maxima are located at $\phi^\pm_{max} = \pm e^{-(\sqrt{5}+1)/2}$, with $V(\phi^\pm_{max})=e^{-(\sqrt{5}+1)}(\sqrt{5}+1)/4$. The minimum at the origin is special, because $V_{\phi\phi}(0)\to\infty$. So, we investigate the solutions that connects it to its neighbor zero. Since the potential presents the $Z_2$ symmetry, we only investigate the interval $\phi\in[0,e^{-1}]$.

\begin{figure}[htb!]
\centering
\includegraphics[width=6cm]{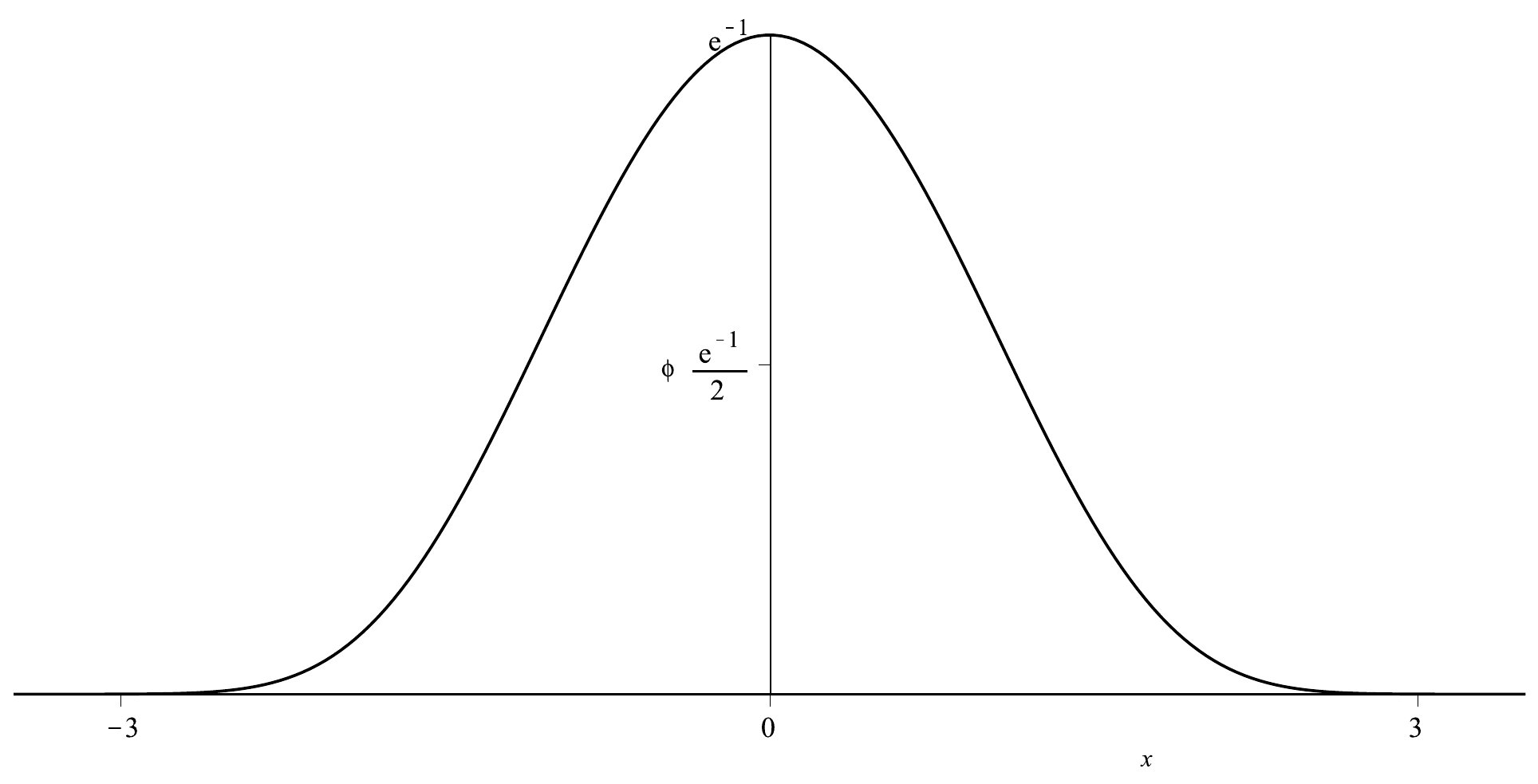}
\caption{The solution in Eq.~\eqref{solde}. Even though it decays faster than the standard solution in Eq.~\eqref{solphi3}, it is not compact.}
\label{fig6}
\end{figure}

In this case, the field equation \eqref{eom} becomes
\be
\phi^{\prime\prime} = \phi\left(\ln^2(\phi) + \ln(\phi) -1\right),
\ee
which admits the following solution
\be\label{solde}
\phi(x) = e^{-\cosh(x)},
\ee
which is a lump that decays as $\phi(x\to\pm\infty) \to e^{-e^{\pm x}/2}$. Here, we observe a novel feature. Notice that, even though $V_{\phi\phi}(0)\to\infty$, which happens for compact lumps in Ref.~\cite{complump}, the above solution decays much faster than the standard one, but it is not compact. In Fig.~\ref{fig6}, we display this solution. As one knows, among the aforementioned applications, lumps provide the basic tooling to study Q-balls. In Ref.~\cite{compqballs}, it was introduced, with the inspiration of Ref.~\cite{complump}, compact Q-balls, which presents both the solution and the energy density compactified. The compactification of these quantities modifies the usual behavior of the quantum mechanical stability of Q-balls, since compact Q-balls never decay in free particles. Specifically, this happens due to the behavior $V_{\phi\phi}(0)\to\infty$. Therefore, models of the type described by the potential in Eq.~\eqref{vde} are of interest in the Q-ball scenario, where they may lead to non compact structures that never decay in free particles, so they behave as the compact ones regarding the stability.

\begin{figure}[t!]
\centering
\includegraphics[width=4.2cm]{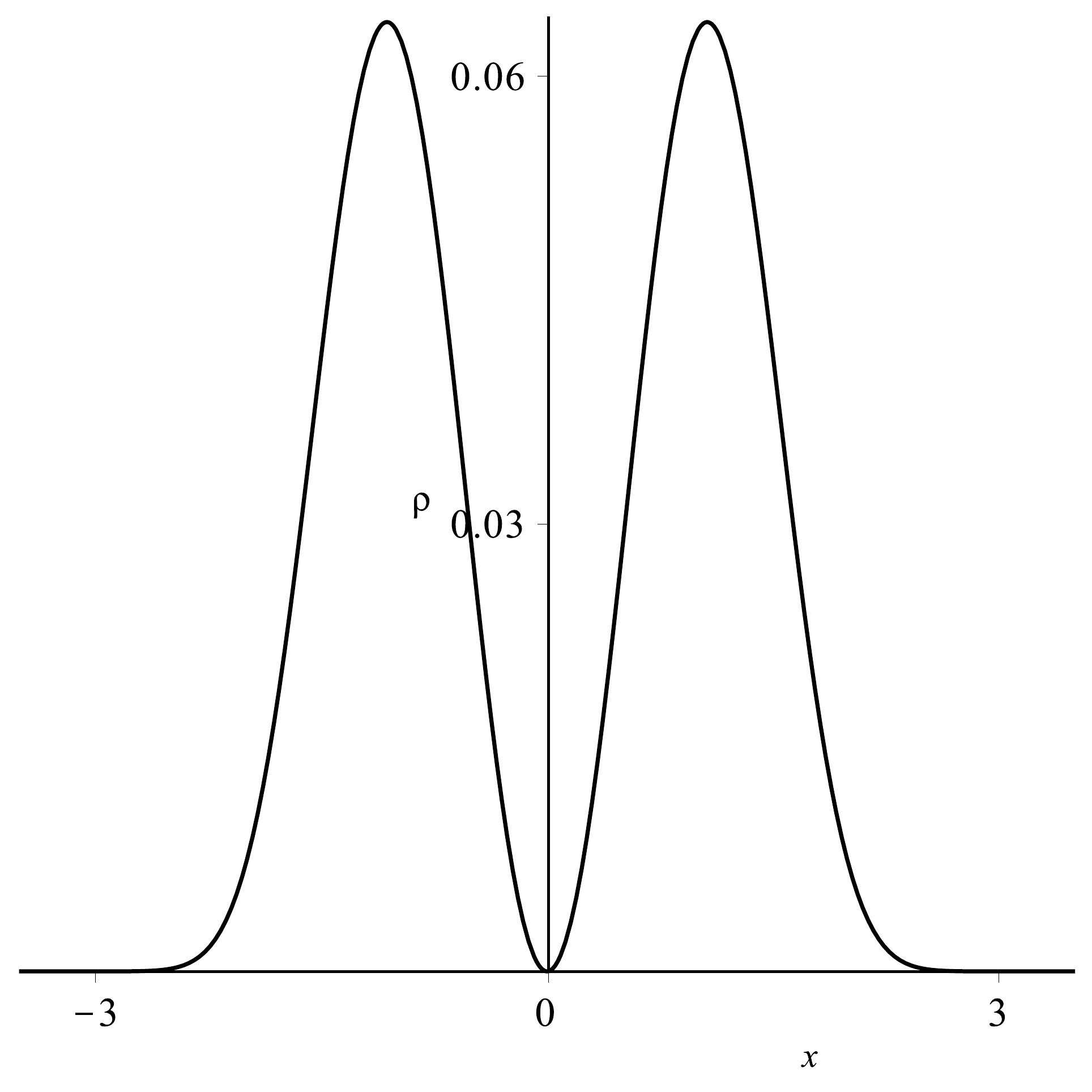}
\includegraphics[width=4.2cm]{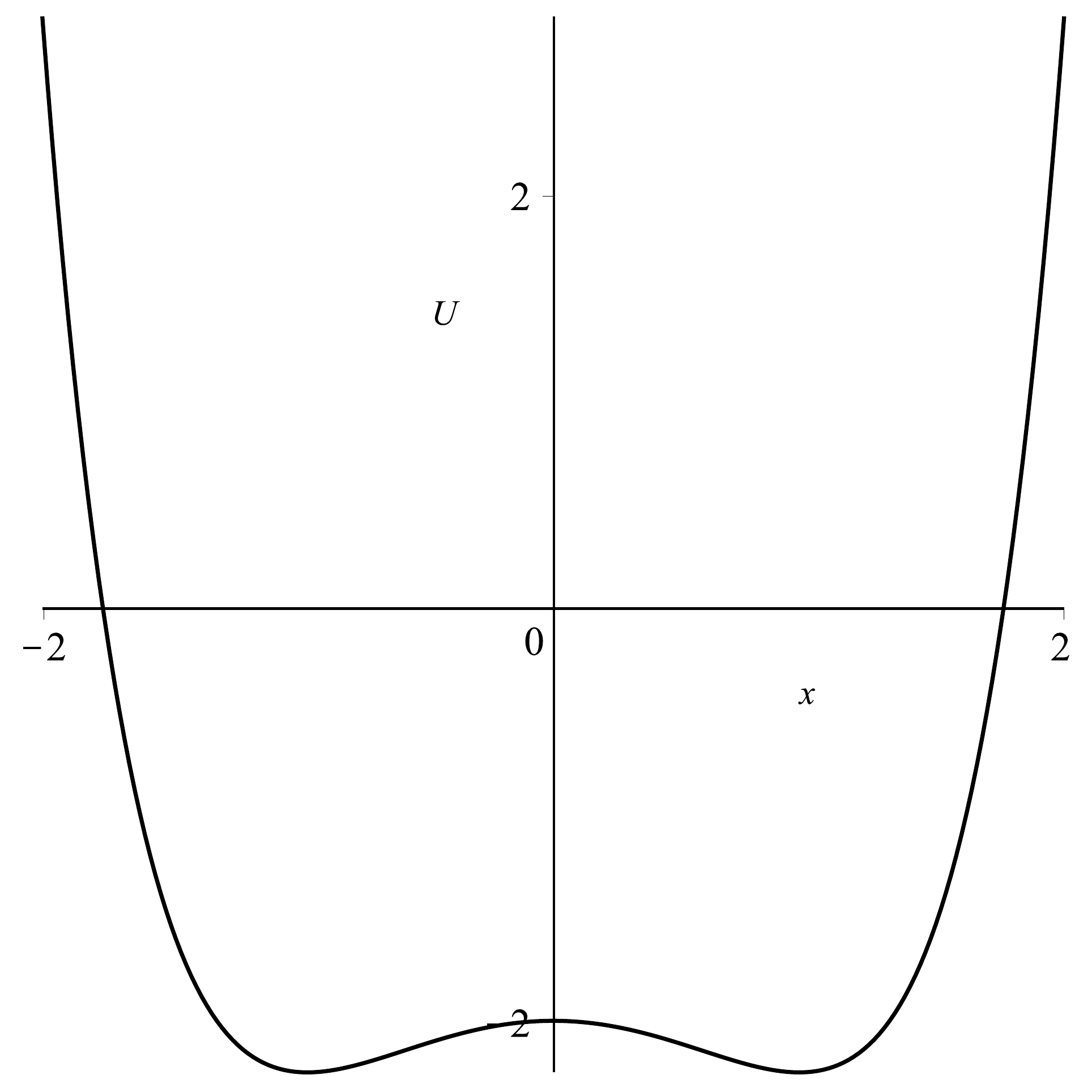}
\caption{The energy density in Eq.~\eqref{rhode} (left) and the stability potential in Eq.~\eqref{ude} (right).}
\label{fig7}
\end{figure}

The energy density can be obtained from Eq.~\eqref{rho}, which leads to
\be\label{rhode}
\rho(x) = e^{-2\cosh(x)}\sinh^2(x).
\ee
It is plotted in Fig.~\ref{fig7}. By integrating it, we get the energy $E=0.139$. We also investigate the linear stability as in the previous examples. In this case, the stability potential in Eq.~\eqref{sta} takes the form
\be\label{ude}
U(x) = \cosh(x)\left(\cosh(x) - 3\right).
\ee
We can see it in Fig.~\ref{fig7}; notice that it goes to infinity asymptotically. Since the zero mode is $\eta_0(x) = e^{-\cosh(x)}\sinh(x)$, which presents a node, the above potential admits a negative bound state. Thus, the lump in Eq.~\eqref{solde} is unstable under small fluctuations.

\section{Outlook}
In this paper, we have investigated lumplike solutions in scalar field models in two flat spacetime dimensions. We have presented some of their general properties and have shown that these structures do not attain a topological charge and are unstable under small fluctuations. As we discussed, the standard lump presents an exponential decay, which is a consequence of a nonvanishing $V_{\phi\phi}$ calculated at the minimum in consideration.

We then introduced two new models that presents analytical solutions with new features. In the first example, we studied a polynomial potential that gives rise to lumps with polynomial decay. Due to this, these solutions presents a large tail if one compares it to the standard case. The other illustration dealt with a potential that presents a logarithmic term. In this case, we have obtained a solution that engender a double exponential decay. It is worth to emphasize that, even though $V_{\phi\phi}\to\infty$ at the minimum, the lump is not compact as in Ref.~\cite{complump}.

The aforementioned models are unstable under small fluctuations, since their corresponding zero mode in the stability equation presents a node. Nevertheless, one can try to work this around by enlarging the model with the addition of charged fermions. Other possibility is to modify the starting model and consider it to be described by a complex scalar field, which allows for the presence of Q-balls. Moreover, one can investigate the presence of lumplike excitations with the aforementioned new features as axions \cite{R} and as tachyonic branes in a warped geometry in the curved spacetime scenario \cite{B,BJ}. Some of these issues are currently under consideration and will be reported elsewhere.

\acknowledgments{We would like to thank Dionisio Bazeia and Roberto Menezes for the discussions that have contributed to this work. We would also like to acknowledge the Brazilian agencies CNPq (research project 140735/2015-1) and CAPES (PDSE process 88881.189443/2018-01) for the financial support.}


\end{document}